\documentclass[preprint]{aastex}
\usepackage{psfig}

\begin{document}

\title{The Rise Times of High and Low Redshift Type Ia Supernovae
are Consistent}

\author{Greg Aldering, Robert Knop, and Peter Nugent}

\affil{Lawrence Berkeley National Laboratory, M.S. 50-232, 1 Cyclotron
Road, Berkeley, CA 94720}

\email{galdering, raknop, penugent@lbl.gov}

\begin{abstract}

We present a self-consistent comparison of the rise times for low--
and high--redshift Type Ia supernovae. Following previous studies, the
early light curve is modeled using a $t^2$ law, which is then mated
with a modified Leibundgut template light curve. The best-fit $t^2$
law is determined for ensemble samples of low-- and high--redshift
supernovae by fitting simultaneously for all light-curve parameters
for all supernovae in each sample.  Our method fully accounts for the
non-negligible covariance amongst the light-curve fitting parameters,
which previous analyses have neglected. Contrary to
\citet{riess_evol99}, we find fair to good agreement between the rise
times of the low-- and high--redshift Type Ia supernovae.  The
uncertainty in the rise time of the high-redshift Type Ia supernovae
is presently quite large (roughly $\pm 1.2$ days statistical), making
any search for evidence of evolution based on a comparison of rise
times premature.  Furthermore, systematic effects on rise-time
determinations from the high-redshift observations, due to the form of
the {\it late}-time light curve and the manner in which the light
curves of these supernovae were sampled, can bias the high-redshift
rise-time determinations by up to $^{+3.6}_{-1.9}$ days under extreme
situations.  The peak brightnesses --- used for cosmology --- do not
suffer any significant bias, nor any significant increase in
uncertainty.

\end{abstract}


\section{Introduction}

Two independent research groups have presented compelling evidence for
an accelerating universe from the observation of high-redshift Type Ia
supernovae (SNe~Ia) \citep{42SNe_98,riess_scoop98}.  These findings
have such important ramifications for cosmology that every effort must
be made to thoroughly test the calibrated standard candles on which
they are based.  Indeed, these groups, and others, are pursuing
additional observations at both high- and low-redshift to confirm
these results. There are programs in place aimed at reducing the
statistical errors, testing systematic errors, limiting the amount of
absorption due to grey dust \citep{aguirre_99}, and searching for
signs of evolution as a function of redshift in SNe~Ia.

Recently \citet{riess_evol99} attempted to examine the question of
whether the rise times of SN~Ia evolve. They used new low-redshift
SNe~Ia light-curve photometry from \citet{riess_rise99} to compare the
mean rise time of these SNe~Ia to a preliminary rise time for
high-redshift SNe~Ia given in a conference abstract by
\citet{don_baas98} and based on a {\it composite} light curve derived
from Supernova Cosmology Project (SCP) observations.  Riess {\it et
al.} noted a 5.8-$\sigma$ difference between the rise times from the
low-redshift data and from the \citet{don_baas98} preliminary analysis
of high-redshift data, with the high-redshift supernovae having shorter
rise times by 2.4 days. Based on this result, they suggested the
possibility that SNe~Ia undergo sufficient evolution to account for
what has been interpreted as evidence for an accelerating universe.

In what follows, we address major shortcomings of these earlier analyses
which fundamentally alter the conclusion of \citet{riess_evol99}.
Specifically, the analysis method used in \citet{don_baas98} to produce
a high-redshift rise-time estimate is very different than that used to
produce the low-redshift rise-time estimate of \citet{riess_evol99}.
Furthermore, both analyses neglected correlated uncertainties in the
light-curve fit parameters, and amongst the light-curve data points, so
neither of these analyses is complete. We also examine, in $\S3$, the
role of light-curve sampling differences between the low-redshift and
high-redshift SN~Ia observations and how they can conspire with
systematic deviations from the fitted reference template --- seen for
normal SNe~Ia --- to shift the inferred rise time. In $\S4$ we briefly
discuss the (small) impact on the cosmological application of SNe~Ia
resulting from light-curve variations. We conclude in $\S5$ with a
summary of our results and a discussion intended to help guide future
work on the question of whether SNe~Ia evolve.

\section{Statistical Analysis of SNe~Ia Rise Times}

\subsection{Description of the Problem}

\begin{figure}[p]
\psfig{file=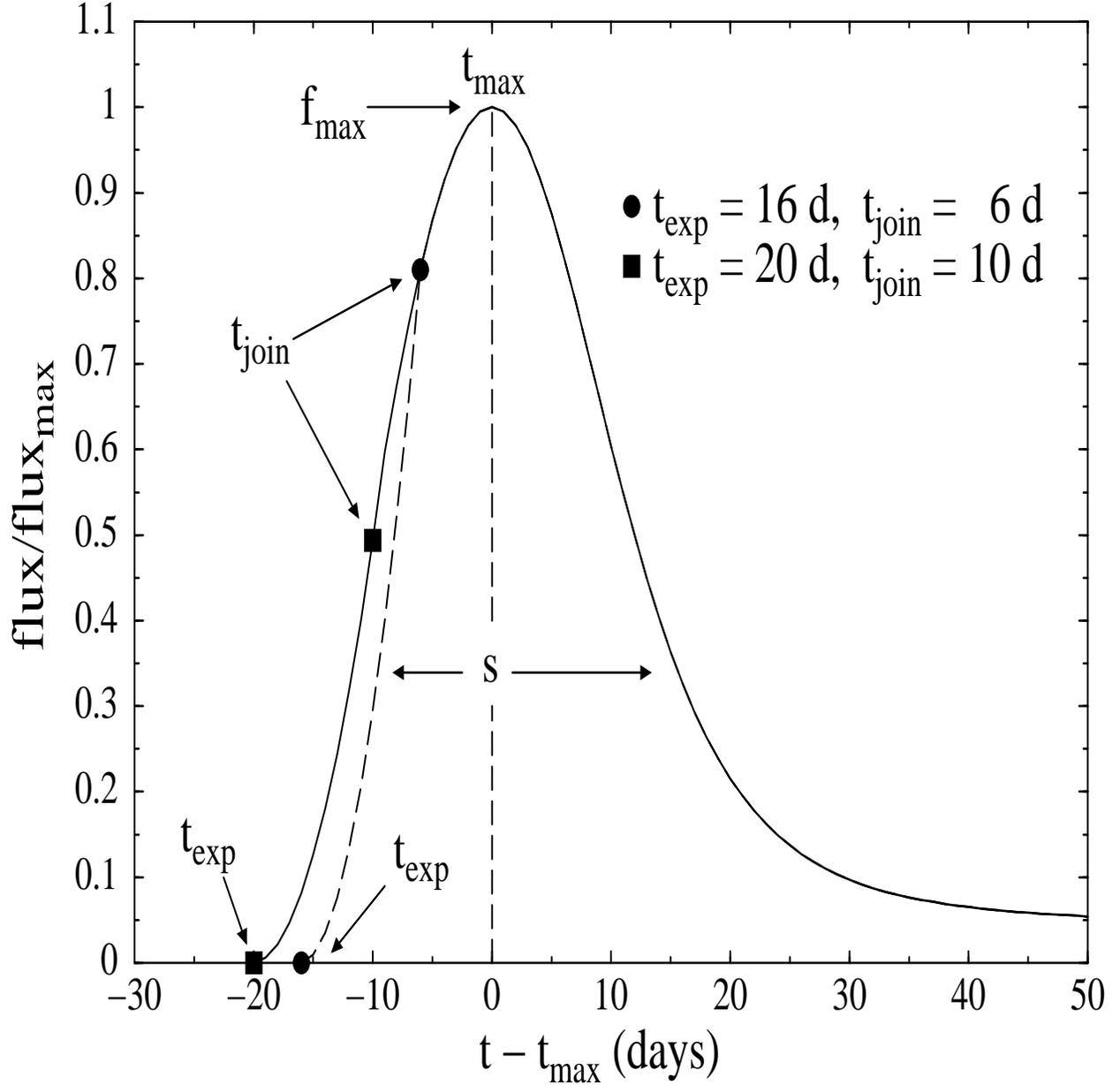,height=6.5in,width=6.5in,angle=270}
\figcaption[templ.eps]{The modified Leibundgut template used by the
SCP \citep{42SNe_98,perl97}, with the standard light-curve fitting
parameters, $f_{max}$, $t_{max}$, and $s$ labeled. Also shown are two
examples from the grid of $t^2$ laws mated to the modified Leibundgut
template used to perform the statistical fits discussed in the
text. The $t^2$ law is parameterized using an explosion day,
$t_{exp}$, and a date, $t_{join}$, at which the $t^2$ law mates with
the modified Leibundgut template.\label{templ}}
\end{figure}

Figure~\ref{templ} illustrates the full SN~Ia template, $\psi(t)$,
normally used by the SCP, which is a modified version of the
Leibundgut template \citep{brunophd,perl97}. The light-curve fitting
parameters are the peak flux, $f_{max}$, time of maximum, $t_{max}$,
and light curve stretch, $s$. (Note that all time dependent quantities
refer to the rest frame of the supernova.) \citet{gerson_baas98} has
demonstrated the remarkable fact that the stretch method applies to
the rising portion of SN~Ia light curves as well as it applies to the
declining portion (up to +25 days after maximum) to better than 2\% of
the peak flux. This has been confirmed for nearby SNe~Ia by
\citet{riess_rise99}. One can represent the flux light curve, $f(t)$,
as follows: $$ f(t) = f_{max} \psi((t - t_{max})/s)$$ This approach
works well in the $U-$, $B-$ and $V-$bands over the range $-20\ {\rm
days} < t - t_{max} < +25 \ {\rm days}$ (see both \citet{42SNe_98} and
\citet{perl97} for a full explanation of the use of this approach).

A meaningful comparison of rise times for low-- and high--redshift
supernovae requires that both datasets be fit with the same template,
and that the fits be performed in a manner which fully accounts for
the covariance between the light-curve fitting parameters and the
calculated rise time.  For the high-redshift data, accounting for
covariance in the light-curve fitting parameters is especially
important since the uncertainties on individual data points are
relatively large.  Such uncertainties allow the fitted date of maximum
light, $t_{max}$, the peak brightness, $f_{max}$, and the light-curve
width, $s$, to be changed in compensating ways to yield similarly good
fits.  Thus, these parameters are correlated, and since determination
of the rise time or explosion date, $t_{exp}$, involves both $s$ and
$t_{max}$, it is incorrect to fit for these parameters while holding
$s$ and $t_{max}$ fixed.

Take for example the case where the fitted value of $t_{max}$,
$t_{max}^\prime$, is too early by 1~day. The fitted value of $s$,
$s^\prime$, will suffer a compensating increase by roughly $1/15$ in
an effort to fit the data on the fast-declining, well-sampled portion
of the light curve at $+10 < t - t_{max} < +20\ {\rm days}$.  The
effective stretch-corrected epoch, $t_s = (t - t_{max})/s$, of a point
nominally at $t-t_{max} = -20$~days and for $s = 1$ would be incorrect
by:
$$\Delta t_s = (t - t_{max})/s - (t - t_{max}^\prime)/s^\prime$$
$$ = \frac{-20}{1.00} - \frac{-19}{1.07} = -2.2\ {\rm days}.$$
Likewise, if $t_{max}^\prime$ were 1~day after $t_{max}$, $s^\prime$
would be smaller than the true $s$, changing $\Delta t_s$ by roughly
$+2.5$~days. This is the principal mechanism by which uncertainties in
the light-curve fit parameters propagate into increased uncertainty in
SN~Ia rise times. (Our Monte Carlo simulations in \S3 bear this out.)
If the uncertainties in $t_{max}$ and $s$ had simply been propagated
as if they were independent, the assigned uncertainty would be 1.7
days, and the correlated nature of the uncertainties would be lost.
It is true that a point at $t-t_{max} = -20$~days may also play some
role in constraining $s$. However, for the datasets considered here
the observations on the rising portion of the light curves are
generally much less certain than those on the declining portion.

The analyses presented in \citet{don_baas98} and \citet{gerson_baas98}
were designed to test the efficacy of the stretch technique when
applied to the rising portion of the light curves of high-redshift
supernovae, and to attempt to improve that portion of the SCP light
curve template. The high-redshift data from the SCP were aligned to
stretch-corrected epochs, $t_s$, using $t_{max}$ and $s$ for each
supernova determined from individual light-curve fits without exclusion
of data from any light-curve epoch. Then a $t^2$ rise-time model was
fit to the ensemble pre-max data, with the final result quoted for
$t^2$ fits covering rest-frame epochs $-21$ to $-10$ days with respect
to $t_{max}$. None of the uncertainty due to the light-curve fitting
parameters was propagated into the final quoted rise-time uncertainty 
\citep{gerson_privcom}.  The resulting $t^2$ fit was then used to
develop a revised template, and the individual SNe~Ia light curves were
then re-fit to this revised template.

\citet{riess_evol99} analyzed the low-redshift data very differently:
they aligned their low-redshift data using $t_{max}$ and $s$ for each
supernova as in the preliminary high-redshift analysis, but they used
only data from $-10$ to $+35$ days to fit the light curves. After
aligning the light curves, a $t^2$ rise-time model was fit to the
ensemble pre-max data, with the final result quoted for $t^2$ fits
covering rest-frame epochs $-23$ to $-10$ days. Following
\citet{riess_rise99}, the uncertainty in $t_{max}$ and $s$ was
accounted for in \citet{riess_evol99} by increasing the uncertainties
on the stretch-corrected light-curve photometry points. The modest
contribution due to correlated uncertainties was not included.


Both of these studies fixed $t_{max}$ and $s$ for the individual SNe~Ia
before fitting the $t^2$ model from which explosion dates were
inferred. They propagated the uncertainties in the light-curve fits
parameters in an incomplete and approximate way. Since this approach
does not allow each individual supernova's light-curve fit parameters,
$f_{max}$, $t_{max}$, and $s$, to adjust to give the best fit as
different rise times are tested, the uncertainties quoted in these
studies are likely to be underestimates. In addition, since the two
studies fit to different time intervals of data, a comparison of the
central values may not be self-consistent.

\subsection{Fitting Method}

The most assumption-free means of accounting for how the uncertainties
in the fits to individual SNe~Ia light curves affect the value and
uncertainty of the rise time is to explicitly test various rise times
to see how well the SNe~Ia are able to adjust to give fits of similar
quality. This is more accurate than, and avoids difficulties
associated with, attempting to propagate uncertainties based on the
covariance matrix determined at the best-fit value when dealing with
complex parameter probability spaces, such as those which occur for
some SNe~Ia light curves dealt with here.  This approach requires that
a family of templates with different rise times be defined and fit to
the entire photometric dataset for each SN~Ia.

Unfortunately, at present, very little light-curve data are available
for determining a suitable early-epoch template for a SN~Ia. Therefore
we have constructed a grid of templates consisting of $t^2$ models
starting with zero flux at an explosion epoch, $t_{exp}$, and joined
to the modified Leibundgut template at epoch, $t_{join}$.  A $t^2$
model can be justified under the conditions of uniform expansion and
constant effective temperature from simple physics (see also \citet{arnett82}).
Two examples from this family of $t^2$-model, $t_{exp}$, $t_{join}$
templates are shown in Figure~\ref{templ}, with the epochs $t_{exp}$
and $t_{join}$ labeled.  These can be compared to the modified
Leibundgut template, which is known to be a reasonable approximation
to the light curves of many SNe~Ia (with the timescale stretched or
contracted).

Note that the use of $t_{exp}$, $t_{join}$ to describe the early-epoch
light curve is simply a reparameterization of the $\alpha, t_{exp}^2$
models (i.e., $f(t) = \alpha (t - t_{exp})^2$) used in previous
studies, with the added constraint of continuity where the
$\alpha, t_{exp}^2$ model ends and the modified Leibundgut template
begins. \cite{riess_evol99} did not impose a continuity constraint
since the fitting to the early stretch-corrected light curve with an
$\alpha, t_{exp}^2$ model was performed {\it after} (some portion) of the
original light curve was fit with another template. \citet{don_baas98}
and \citet{gerson_baas98} have an implicit continuity constraint in
that they mated their best-fit $t_{exp}^2$ model to the remainder of their
light curve when constructing each new template. In this paper the fit
for the rise time and the overall light-curve parameters is performed
simultaneously.

An added benefit of our parameterization is that $t_{exp}$, $t_{join}$
are more nearly orthogonal than $\alpha, t_{exp}^2$. This is because
the already-established modified Leibundgut template provides a strong
constraint on the amplitude of a $\alpha$, $t_{exp}^2$ model at the
point it crosses the modified Leibundgut template. $\alpha$ simply
adjusts itself to satisfy this constraint as $t_{exp}$ is changed. (This
leads to the narrow, but strongly tilted, confidence regions in Figure~1 of
\citet{riess_evol99}).

The fitting method we use integrates the probability [$P \propto
exp(-\chi^2/2)$; see Eq.~28.22 in \citet{pdb_98} ] over the parameters
$f_{max}$, $t_{max}$, and $s$ separately for each supernova, at each
value of {$t_{exp}$, $t_{join}$}. The fits are performed in flux
(rather than magnitudes); this allows the use of non-detections, these
being the principal source of early-epoch data for the
\citet{42SNe_98} high-redshift supernovae.  Two alternative methods
are used to perform the integrations over $f_{max}$, $t_{max}$, and
$s$. In the first, the integral over $f_{max}$ is performed
analytically and the subsequent integration over $t_{max}$, and $s$
uses the adaptive integration algorithm of \citet{acm698}. The second
method uses a grid of $\Delta f_{max} = 0.01$, $\Delta t_{max} =
0.1$~days and $\Delta s = 0.01$ centered on the averages of the best
fit values over $t_{exp}$ for the high-redshift SNe~Ia. To account for
the tightly constrained parameters of the low-redshift SNe~Ia a hybrid
technique is used in which the integral over $f_{max}$ is done
analytically and a grid of $\Delta t_{max} = 0.01$~days and $\Delta s
= 0.0005$ is used to integrate over $t_{max}$ and $s$. In each, the
limits are chosen such that the probabilities are negligible at the
boundaries.  We find excellent agreement between each of these
methods.

The end product is a map of probability over {$t_{exp}$, $t_{join}$}
for each supernova. These probability maps are then multiplied, then
renormalized, for an ensemble of supernovae, e.g., the high-redshift
supernovae from \citet{42SNe_98} or the low-redshift supernovae of
\citet{riess_evol99}, to determine the joint probability distribution
function over {$t_{exp}$, $t_{join}$}, $P(t_{exp}, t_{join})$; or
after normalizing over $t_{exp}$ for each $t_{join}$, the 
conditional probability distribution function for $t_{exp}$ given
$t_{join}$, $P(t_{exp} | t_{join})$.

\subsection{Supernova Light-Curve Samples}

What we will hereafter refer to as the ``low-redshift SNe~Ia'' sample
consists of SN~1990N, SN~1994D, SN~1996bo, SN~1996bv, SN~1996by,
SN~1997bq, SN~1998aq, SN~1998bu, and SN~1998ef, for which early-epoch
light-curve photometry transformed to $B$-band from unfiltered CCD
images has been reported by \citet{riess_rise99}. The early-epoch
photometry was supplemented with data from \citet{lira98, patat94d,
meik94d, riess_data99, sunetal99, jha99_98bu, riess_rise99} to produce
full $B$-band light curves extending over peak and beyond.
\citet{riess_rise99} reports four early-epoch light-curve points (one
an upper limit) for SN~1998dh, however we were unable to include this
supernova since the subsequent light-curve photometry was unavailable.

What we will hereafter refer to as the ``high-redshift SNe~Ia'' sample
consists of the 30 SNe~Ia from \citet{42SNe_98} having redshift $0.35
< z < 0.65$, with the exception of SN~1997aj\footnote{SN~1997aj was
excluded due the presence of several highly deviant points in its
light curve (including large deviations within a given night) which
for some combinations of $t_{exp}$ and $t_{join}$ produced fits with
greatly improved values of $\chi^2$, but unacceptably large values of
stretch. Inclusion of SN~1997aj gave longer rise times, in better
agreement with \cite{riess_evol99}, and reduced the rise-time
difference by $\sim$~0.9~days compared to our results in \S2.4.  Thus,
although SN~1997aj was found to reinforce the findings discussed
below, the most conservative choice was to eliminate this SN.}. As
defined, this sample satisfies the requirements that at least 60\% of
the light in the $R$-band comes from the rest-frame $B$-band and that
at least 60\% of the rest-frame $B$-band light is included in the
$R$-band.  Redshift limits satisfying these conditions were determined
using the $B$-band and $R$-band filter responses given in
\citet{bess90}, along with spectra of normal SNe~Ia as a function of
light-curve epoch constructed by \citet{nuge_kcorr00}.  These
restrictions allow comparison with the low-redshift $B$-band
photometry of \citet{riess_evol99} while minimizing the potential
uncertainties inherent in making large cross-filter
K-corrections. Even so, K-correction uncertainties will be present
for those supernovae near these redshift limits, as well as at very
early times where few spectra are available from which K-corrections
can be calculated.  Note that most of the other eleven supernovae from
\citet{42SNe_98} have complete light curves only in rest-frame
$V$-band or $U$-band, and therefore are unsuitable for determination
of the $B$-band light-curve parameters.  Also note that not all 30
high-redshift SNe~Ia have equal value in determining the rise
time. Only those that were fortuitously caught on the rise in the
reference images of the search run can constrain this region of the
light curve.

\subsection{Results of the Statistical Analysis}

\begin{figure}[p]
\psfig{file=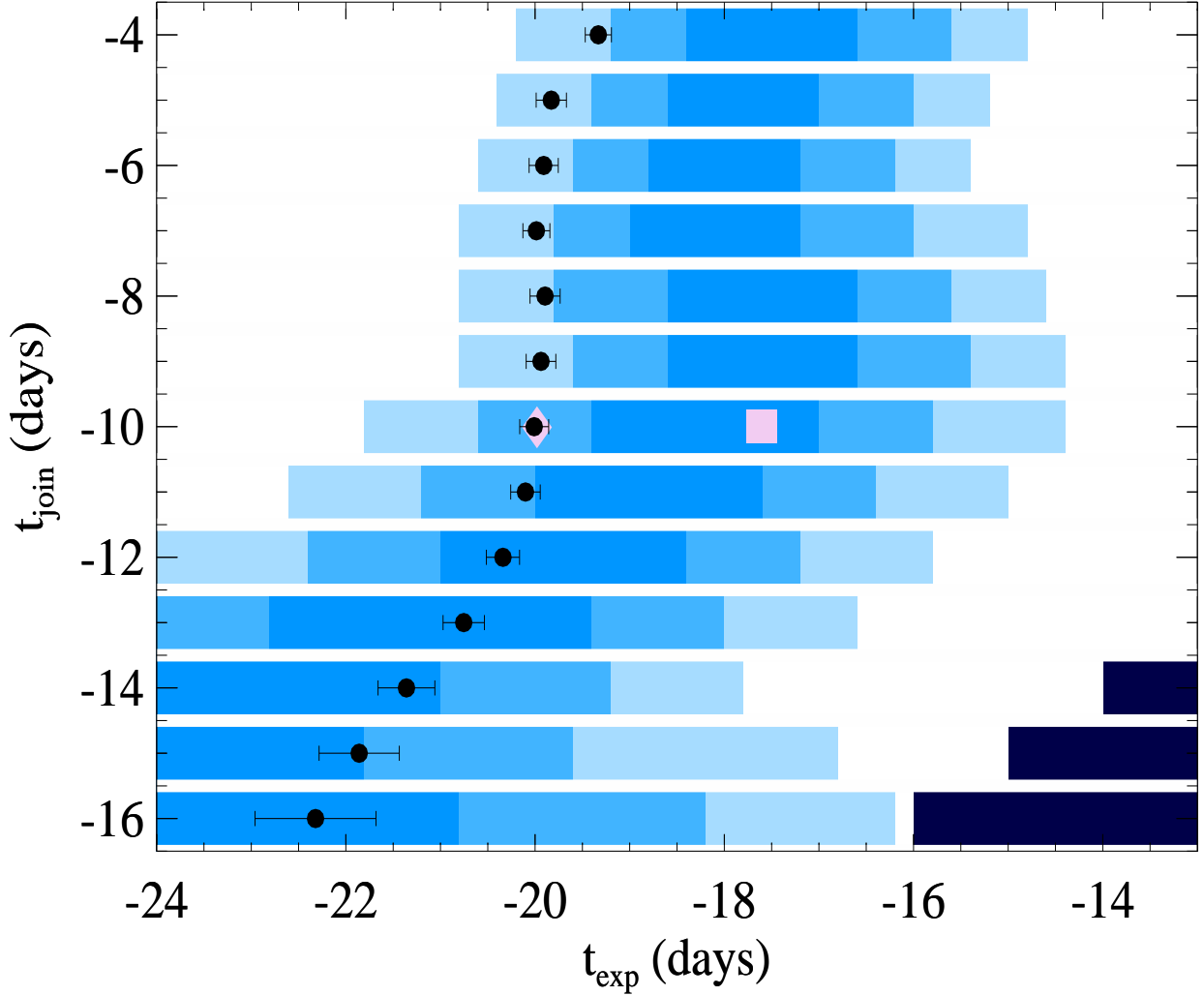,height=5.5in,width=6.5in,angle=90}
\figcaption[conf.eps]{The 1--, 2--, and 3--$\sigma$ (68.3\%, 95.4\%,
and 99.73\%, in progressively lighter shades of blue) confidence
regions for the {\it conditional} probability of $t_{exp}$ given
$t_{join}$, $P(t_{exp} | t_{join})$, for the high-redshift SNe~Ia
sample.  The small solid circles are our results for the
maximum-likelihood $t_{exp}$ for a given $t_{join}$ for the
low-redshift sample. The uncertainties in $t_{exp}$ for these points
vary from 0.14~days at $t_{join}=-4$ to 0.64~days at $t_{join}=-16$.
For most values of $t_{join}$, the values of $t_{exp}$ for the
low-redshift and high-redshift samples are in fair to good agreement.
The large solid diamond represents the best-fit $t_{exp}$ for the
low-redshift supernovae found by \citet{riess_evol99}, for $t_{join} =
-10$ days. The large solid square is the best-fit $t_{exp}$ for the
high-redshift supernovae from the preliminary work of
\citet{don_baas98} and \citet{gerson_baas98}.  The best-fit values
found in this work for $t_{join} = -10$ days are in good agreement with
these previous studies, especially given the larger uncertainties
(mainly for the high-redshift SNe~Ia sample) we find with our
analysis.  Note that the region $t_{exp} < t_{join}$ (shaded in black)
is physically excluded by the requirement that a supernova light curve
be a single-valued function of time.
\label{conf}}
\end{figure}

\begin{figure}[p]
\psfig{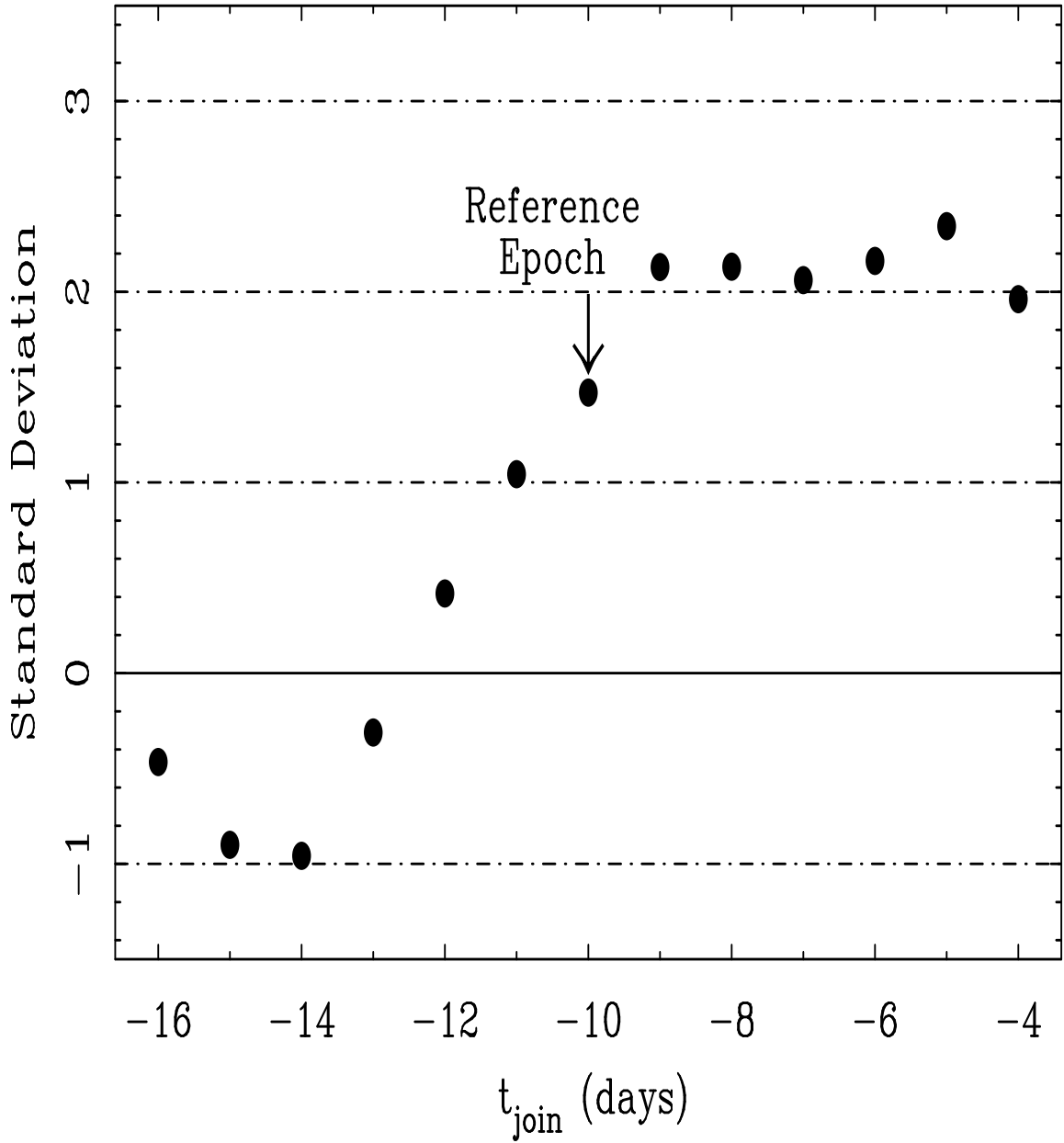}
\figcaption[dev_join.eps]{The equivalent number of Gaussian standard
deviations by which the best-fit $t_{exp}$ for the low-redshift SNe~Ia
sample differs from that for the high-redshift SNe~Ia sample, for each
value of $t_{join}$. This calculation accounts for the uncertainties
in $t_{exp}$ from both datasets. The difference of 1.5$\sigma$\ at
$t_{join} = -10$ days is shown as a point of reference, as the studies of
\citet{riess_evol99, don_baas98, gerson_baas98} have fit $t^2$ laws up
through -10 days (to data that was pre-aligned by light-curve epoch and
stretch). Note that for any value of $t_{join}$ the differences are much
less than the 5.8 standard deviations found by \citet{riess_evol99}.
\label{dev_join}}
\end{figure}

Templates were generated for $-29.9 < t_{exp} < -10.1$ days, in steps
of 0.2~days, and for $-20 < t_{join} < -4$ days, in 1 day steps.
Fitted templates were required to have $t_{exp}$ earlier than
$t_{join}$.  Figure~\ref{conf} presents the results of these fits;
shown are the 1--, 2--, and 3--$\sigma$ confidence regions for the
conditional probability, $P(t_{exp} | t_{join})$, for the
high-redshift SNe~Ia sample.  Also shown are points which mark the
most probable value of $t_{exp}$ at each $t_{join}$ for the
low-redshift SNe~Ia sample.  Figure~\ref{dev_join} distills the
$t_{exp}$ differences taken from Figure~\ref{conf} into equivalent
Gaussian standard deviations for the difference in $t_{exp}$ between
the high-redshift and low-redshift SNe~Ia samples.  These plots
demonstrate that for $t_{join} < -10$, the high-redshift and
low-redshift SNe~Ia samples agree at the 1-$\sigma$ level or better.
For $t_{join}$ less than $\sim -15$~days, the high-redshift SNe~Ia
sample is unable to place meaningful constraints on $t_{exp}$.

The rise-time value quoted in \citet{riess_evol99} of $t_{exp} =
-19.98\pm0.15$ --- compared to $t_{exp} = -20.08\pm0.19$ (statistical)
obtained from our analysis --- was determined for $t_{join} = -10$ days,
and is plotted in Figure~\ref{conf}. Even at this reference epoch the
disagreement between the high-redshift and low-redshift SNe~Ia samples
is only 1.5-$\sigma$, {\it not} the 5.8-$\sigma$ difference found by
\citet{riess_evol99}.  The value of $t_{exp} = -17.6\pm0.4$ days given
in preliminary analysis of the high-redshift sample by
\citet{don_baas98} is also plotted in Figure~\ref{conf}.  As
Figure~\ref{conf} shows, the main difference between our finding and
that of \citet{riess_evol99} lies in different best-fit values and
larger uncertainties for the high-redshift SNe~Ia sample (differing by
$-0.7$\ days at $t_{join} = -10$ days). The uncertainties are larger,
especially for the high-redshift SNe~Ia sample, when uncertainties in
the light-curve fit parameters, $f_{max}$, $t_{max}$, $s$ (and to a
lesser extent amongst the photometry points) are fully taken into
account.  These larger uncertainties come about because the individual
SNe~Ia are given the proper freedom to adjust to templates away from
the global best-fit template.  Previous analyses have artificially
suppressed this freedom, and have therefore underestimated the
uncertainty on $t_{exp}$.

Given the large uncertainty in $t_{exp}$, potential perturbations from
the systematic effects discussed in the next section, and the fair to
good agreement in $t_{exp}$ between the low- and high-redshift SNe~Ia
for reasonable values of $t_{join}$, we consider a detailed analysis of
the best $t_{join}$ unwarranted.  \citet{riess_rise99} found that
$\chi^2$ per degree of freedom deteriorated for their fits for
$t_{join} > -8$ days, indicating that the simple $\alpha, t_{exp}^2$
model is not appropriate later than $-8$~days. A cursory examination of
the joint probability, $P(t_{exp},t_{join})$, for our fits showed that
the low-redshift SNe~Ia sample prefers $t_{join} \sim -8$ days, where
our analysis finds a modest disagreement between the low-redshift and
high-redshift supernovae. However, the early low-redshift SNe~Ia
observations prefer a slightly different $t_{join}$;
$P(t_{exp},t_{join})$ based on observations having $t-t_{max} < -6$
days gives a preferred $t_{join} \sim -12$ days, where high- and
low-redshift rise times agree quite well.  This mild tension within the
low-redshift SNe~Ia sample with regard to the preferred $t_{join}$ is
somewhat less than the 2--$\sigma$ level.  A similar, but weaker,
situation is found for the high-redshift SNe~Ia sample.  This is not a
complete surprise; as the following section demonstrates, there are
systematic variations in the late-time light-curve behavior of SNe~Ia
(such as SN~1994D from the low-redshift SNe~Ia sample) which can affect
the preferred rise time. Furthermore, a best fit value of $t_{join}$
depends not only on the rise-time behavior, but also the accuracy of
the modified Leibundgut template for $t_{join} \le t - t_{max} \le -4$
days (the latest $t_{join}$ tested).  The relative probabilites at
different values of $t_{join}$ include a contribution from the $\alpha,
t_{exp}^2$ model for $t<t_{join}$ and from the Leibundgut template for
$t>t_{join}$.  Because the parameters ($t_{max}$ and $s$) for the
modified Leibundgut template are driven largely by points with $t> -4$
days, any early-time mismatch between the modified Leibundgut template
and the data will degrade the quality of the fit a different amount for
different values of $t_{join}$. This effect should only be of
importance for $t_{join}$ later than about $-10$ days, where the data
are better and where the best-fit  $\alpha, t_{exp}^2$ curves begin to
depart from the (full) modified Leibundgut template. As things stand,
the goodness of fit changes imperceptibly with $t_{join}$ for the
high-redshift SNe~Ia sample.

\section{Systematic Effects}

Given these findings from the statistical analysis it is clear
that there is a reasonable consistency between the rise times of the
high- and low-redshift SNe~Ia. However, it is important to explore the
possibility of systematic effects which have the potential to drive a
fit to another location and/or increase the error bars further. One
such effect arises from application of the stretch relationship when
fitting an observed light curve with a given template.

As mentioned in $\S2.1$, the stretch method works particularly well up
to $t \sim +25$ days past maximum. After this point the light curve of
a SN~Ia leaves the photospheric phase and enters into the nebular
phase. This is marked by a bend in the light curve between +25 and +35
days after maximum light where the rapid drop from peak brightness
slows down into an exponential decline of the light curve. Since this
exponential decline is governed mostly by the radioactive decay of
$^{56}$Co to $^{56}$Fe one would not expect it to ``stretch'' like the
earlier portion of the light curve. In fact, as seen in
\citet{brunophd}, the slopes of the declines are very similar for a
wide range of SNe~Ia light-curve widths. This highlights one of the
current limitations of the stretch method; the entire template,
regardless of epoch, is stretched to fit the data. This is not just a
problem for the stretch method, but for any of the current SN~Ia
template fitting methods, which all employ a one-to-one correlation
between peak brightness and the shape of the light curve. This is a
small effect compared to the peak flux and the typical photometric
uncertainties in current low- and high-redshift data sets. However, it
is important to consider its effect specifically on the measurement of
the rise time.

The amplitude with respect to peak of the aforementioned exponential
decline varies among SNe~Ia. It turns out that the stretch method can
compensate somewhat for these differing amplitudes, providing better
fits in the $\chi^2$ sense, but at the expense of introducing a possible
bias in $s$. Since the amplitude variations during the exponential
decline become apparent at brightnesses similar to those on the rising
portion of the light curve being studied here, and since the data are
generally much better for the later portion of the light curve, the
{\it late}-time light-curve behavior may bias determination of the
rise time.  The effect of this bias on the template fitting method was
studied via a Monte Carlo simulation, as described below.

\begin{figure}[p]
\psfig{file=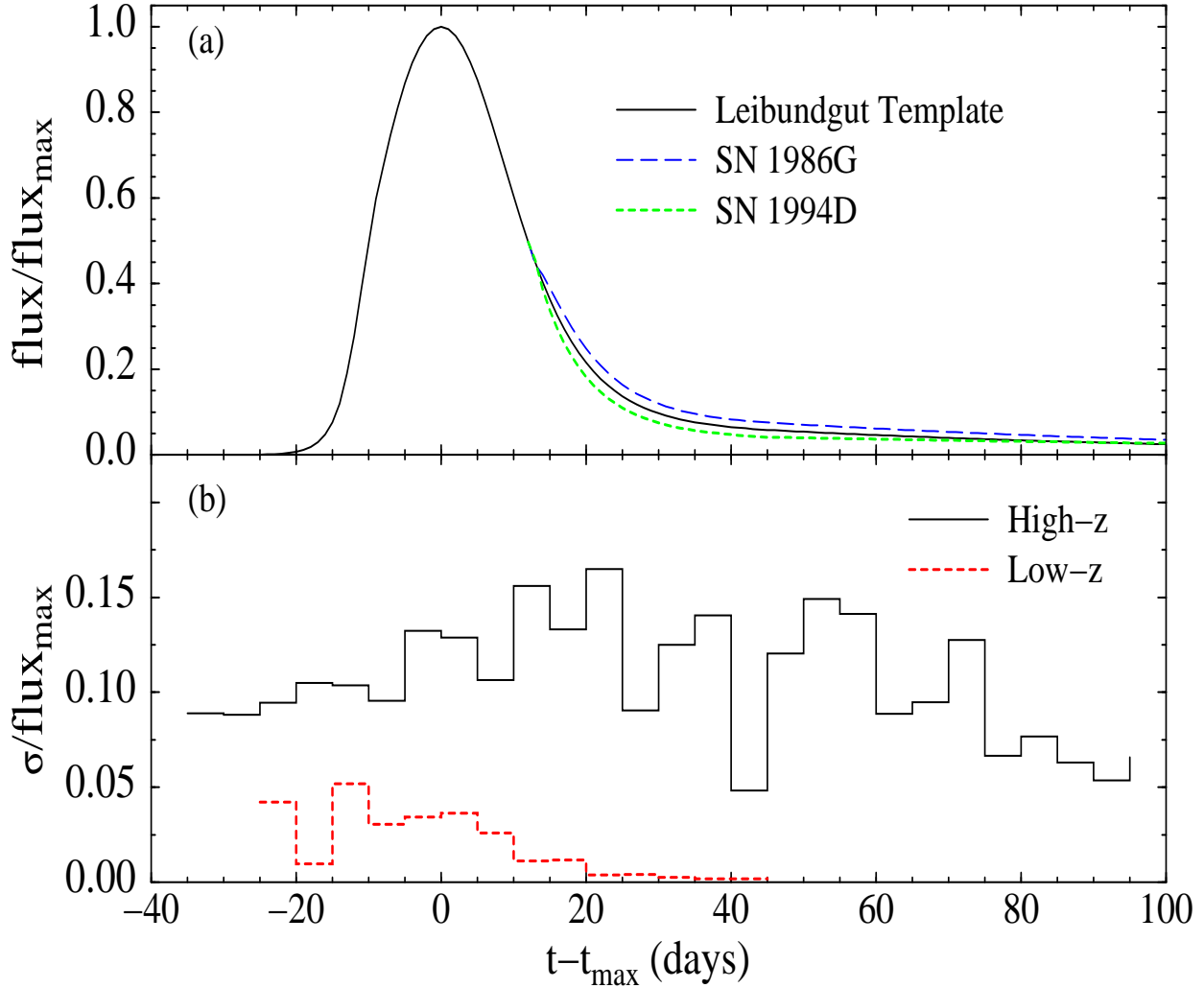,height=5.5in,width=6.5in,angle=270}
\figcaption[samp_sys.eps]{a) The different light-curves shapes used to
study the systematic uncertainty due to temporal sampling and intrinsic
light-curve deviations. The black curve corresponds to the standard
modified Leibundgut template. The blue curve shows the late-time deviation
exhibited by SN~1986G, while the green curve shows the late-time deviation
for SN~1994D. b) The effective normalized ensemble uncertainty as a
function of time for the high-redshift SNe~Ia (black) and low-redshift
SNe~Ia (red) samples.  If the observations from each of these samples had
come from a single high-redshift or low-redshift supernova, this would
reflect how well that light curve is determined. Although the uncertainty
is not strongly dependent on light-curve phase for the high-redshift
SNe~Ia, the scale over which the data can be effectively averaged --- set
by the steepness of the light curve --- leads to a poorer determination
of the early, rising, portion of the light curve. Note also that the
uncertainty is modulated slightly by lunar phase, since the SNe~Ia are
typically discovered shortly before maximum light and one week prior to
new moon, and cannot be observed near full moon.
\label{samp_sys}}
\end{figure}

Figure~\ref{samp_sys}a shows the modified Leibundgut template along
with two other templates derived from the SNe~Ia 1986G and 1994D
\citep{phil86g,meik94d,patat94d}. These supernovae were chosen because,
among those SNe~Ia with good late-time data, they produced the largest
deviations from the modified Leibundgut template in the tail of the
light curve. To produce these templates for the Monte Carlo simulations
the data through day +15 for SNe~1986G and 1994D were adjusted to fit
the modified, unity-stretch, Leibundgut template. The resulting
adjustments were then applied to the data beyond day +15 using the
stretch method.  These adjusted late-time data were fit with a smooth
curve through the bend in the light curve, followed by an exponential
decline. These late-time curves were then mated to the modified,
unity-stretch, Leibundgut template for $t < +15$ days to form complete
templates.

Figure~\ref{samp_sys}b shows the normalized ensemble photometric error
for both the high-redshift and low-redshift SN~Ia samples in 7 day
bins from $ -35 < t - t_{max} < +75$. This indicates how accurately a
light curve would have been measured had all the observations come
from just one supernova. Similarly, provided the stretch method works
sufficiently well, and $f_{max}$, $t_{max}$, and $s$ are known, this
would be the accuracy of a stretch-corrected composite light
curve. Note that the high-redshift data are of consistent quality
through $\sim 50$ days after maximum light, which enables the
high-redshift SNe~Ia sample data to constrain the fit to a template
over a large range in time with nearly equal weight. However, this
makes the high-redshift SNe~Ia data susceptible to a systematic bias
on the rise time due to possible deviations from the stretch fitting
method for $t > 40$ days for deviant light curves like those shown in
Figure~\ref{samp_sys}a.

The Monte Carlo simulation performed to test for such a bias created
simulated light-curve photometry data for the three sets of supernovae
based on the templates seen in Figure~\ref{samp_sys}a. Each set was
comprised of $\sim 100$ different realizations of each of the supernovae in
the high-redshift SNe~Ia sample based on their individual temporal
sampling and associated photometry errors. All of the generated
supernovae were created with the following input parameters: $s =
1.0$, $t_{join} = -10.0$ days, $t_{exp} = -20.0$ days, $t_{max} =
0.0$ and $f_{max}=1.0$.  The resultant light curves produced in each
set were fit with the modified Leibundgut template. $\chi^2$ surfaces
of $t_{exp}$ and $t_{join}$ were created for each of the fits, and
within a set these surfaces were added together to find the global
minimum.

The results of these simulations are given in Table~\ref{sys_tab}.  It
is apparent that given a set of SN~Ia observations like those available
from the high-redshift SNe~Ia sample, a fit for $t_{exp}$ can be biased
by 2---3 days in either direction if all the observed SNe~Ia have
deviant late-type light curves like SN~1986G or SN~1994D.  To allow
direct comparison with Figure~\ref{conf}, these same simulations were used to
determine the best values of $t_{exp}$ for input templates with
$t_{join}$ fixed at $-10.0$ days. For this case we found $t_{exp} =
-19.8$, $-17.5$, and $-22.5$, when the data were simulated using the
Leibundgut, SN~1986G, and SN~1994D templates, respectively. This shows
that systematic errors in $t_{exp}$ are large even when $t_{join}$ is
held fixed.  While the SNe~Ia template light curves used to simulate
the high-redshift data can be thought of as extreme cases, at present
the exact nature and frequency of such deviations at this light-curve
phase is poorly quantified due to a lack of high-quality, well-sampled
observations over peak and through day $\sim$ +60 for nearby
supernovae. Therefore, this result should be taken as a rough upper
limit on the systematic error on $t_{exp}$ due to temporal sampling and
our current limited understanding of how the stretch relationship
should be applied at late times.

\begin{deluxetable}{cccccc}
\tablewidth{0pt}
\tablehead{\colhead{Generating Template} &\colhead {$t_{exp_{in}}$} &
\colhead{$t_{join_{in}}$} & \colhead {$t_{exp_{out}}$} & \colhead
{$t_{join_{out}}$} & $\Delta M_B^{corr}$}
\startdata
Leibundgut & $-20.00$ & $-10.0$ & $-19.76(62)$ & $-8.01(42)$ & $+0.006(011)$\\
SN 1986G   & $-20.00$ & $-10.0$ & $-18.08(59)$ & $-6.81(44)$ & $+0.039(012)$\\
SN 1994D   & $-20.00$ & $-10.0$ & $-23.62(19)$ & $-8.84(20)$ & $-0.020(009)$\\
\enddata
\tablecaption{The resultant shift in $t_{join}$, $t_{exp}$ and $\Delta
M_B^{corr}$ from our Monte Carlo simulations caused by generating
SNe~Ia from the above templates and fitting them with a Leibundgut
template given the sampling and photometric errors in the SCP's data
set. RMS errors in the Monte Carlo simulations are in parens.} 
\label{sys_tab}
\end{deluxetable}

\section{Cosmological Implications}

Assuming all SNe~Ia have rise times similar to that found by
\cite{riess_rise99} from good early-time photometry, a light-curve
template with $t_{exp} \sim -20$~days and $t_{join} \sim -10$~days
might be a better template for use in fitting the light curves
of SNe~Ia at all redshifts. This raises the question of whether such
a change from the modified Leibundgut template to a Riess--like
template would alter the corrected peak magnitudes determined in
\citet{42SNe_98}. In comparing our fits to the high-redshift
SNe~Ia sample using these two alternative templates, we find no
measurable change in the ensemble mean corrected peak magnitudes.
We also find that no individual SN~Ia changed by more than 0.02
magnitudes.

Another obvious question, addressed by the simulations of \S3, is
whether systematic variations in late-time light-curve behavior can
affect the cosmological results of \citet{42SNe_98}. In the last column
of Table~\ref{sys_tab} we list $\Delta M_B^{corr}$, the change in the
ensemble stretch-corrected peak magnitude for each dataset determined
using the stretch-luminosity relation of \citet{42SNe_98}.  These
changes ($-0.039 < \Delta M_B^{corr} < 0.020$) are small, and less than
the systematic biases already considered in \citet{42SNe_98} (0.05
mag).  Given the fact that these simulations represent the most extreme
deviations encountered with our fitting method, we conclude that this
bias has no effect on the determination of the cosmological parameters
from SNe~Ia.

\section{Conclusions \& Discussion}

We find no compelling statistical evidence for a rise-time difference
between nearby and distant SNe~Ia, and therefore no evidence for
evolution of SN~Ia. We do find that for the high-redshift SNe~Ia
sample, temporal sampling coupled with real deviations of SNe~Ia light
curves at late-times could systematically bias the inferred rise time
by 2---3 days. Even if present, these biases cannot dim the peak
magnitudes by more than 0.02~magnitudes nor brighten them by more than
0.04~magnitudes even in the extreme cases that all the distant SNe~Ia
have late-time light curves like SN~1994D or SN~1986G, respectively.
This leaves the cosmological results of \citet{42SNe_98}
unchanged. Due to the large statistical uncertainties and possible
systematic effects, we conclude that the extant photometry of
high-redshift SNe~Ia are in fact poorly suited for placing meaningful
constraints on SN~Ia evolution from their rise times.

If future studies using better early-epoch data (such as that expected
from the {\it SNAP} satellite\footnote{See http://snap.lbl.gov for
information pertaining to the SuperNova Acceleration Probe.}) were to
find significant rise-time differences between nearby and distant
SNe~Ia, would this invalidate the use of SNe~Ia as calibrated standard
candles? This is a very complicated question. However, at least some
models suggest that variations in the early rise-time behavior may be
very sensitive to the spatial distribution of $^{56}$Ni immediately
after the explosion. Such differences would diminish as the SN~Ia
expands and the photosphere recedes, meaning that rise-time variations
wouldn't necessarily translate into differences in peak brightness
\citep{pinto99}. Careful measurement of the rise time and the peak
spectral energy distribution of individual SNe~Ia will have to be
carried out to address this question (see \citet{nugehub95,nugerrat95}
for a full description of the interplay between the rise time and the
spectral energy distribution on the peak brightness of a SN~Ia). It
may even prove possible to use the rise time as an additional
parameter to improve the standardization of SNe~Ia.

We close with some general observations concerning the issue of SN~Ia
evolution. The peak brightnesses of SNe~Ia are determined at some level
by the underlying physical parameters of metallicity and progenitor mass,
whose mean values can be expected to evolve with redshift.  Nonetheless,
there should exist nearby analogs for most distant SNe~Ia since there
is active star formation and a wide range of metallicities within nearby
galaxies \citep{gal_metals,chip_metals}.  The existing empirical relations
between intrinsic luminosity and light-curve shape are able to homogenize
almost all nearby SNe~Ia. This implies that SNe~Ia with some finite (but
as yet poorly quantified) range of metallicities and progenitor masses can
be used as calibrated standard candles.  This forms the basis for using
SNe~Ia at high-redshift to probe the cosmology.  If there is a dominant
population of SNe~Ia whose members are underluminous for their light
curve shape at $z \sim 0.5$, as would be required to explain current
observations in terms of evolution, there should be nearby examples
of these SNe~Ia. Such SNe~Ia are not predominant among nearby SNe~Ia,
as almost all nearby SNe~Ia obey a width-brightness relation. For such
SNe~Ia to predominate at $z \sim 0.5$ while being rare nearby requires
a large reduction in their rate.  Searches for SNe~Ia conducted using
exactly the same CCD-based wide-area blind-search methods used by the
SCP find that the SNe~Ia rate per comoving volume element does not
change significantly between $z < 0.1$ \citep{aldering_nearby}, $z \sim
0.5$ \citep{rate_96, rate_00}, and $z \sim 1.2$ \citep{albinoni_rate}.
For the global rates to stay roughly constant while the rate of such
hypothetical subluminous SNe~Ia changes by an order of magnitude would be
remarkable. For instance, a shift from Pop~II progenitors at $z\sim0.5$
to Pop~I progenitors nearby would result in suppressed rates at $z \sim
0.5$. This is due to the fact that Pop~II stars are a minor contributor
to the luminosity density out to $z \sim 0.5$ \citep{gal_evol}. Quantifying
these arguments is beyond the scope of this paper, so we do not claim they
as yet place a bound on SN~Ia evolution.  However, such arguments should
be borne in mind when weighing the likelihood that the {\it calibrated
peak brightnesses} of SNe~Ia evolve.  These arguments can also provide
a partial basis for rigorous testing of the SN~Ia evolution hypothesis.

\acknowledgments

We would like to thank our colleagues in the Supernova Cosmology Project
for their support and contributions to this work.  In particular, we
thank Gerson Goldhaber, Don Groom, and Saul Perlmutter, for discussions on
the ongoing SCP effort that provides the larger context for the analysis
presented here. The analysis of the data and the Monte Carlo simulations
presented in this paper were performed on the National Energy Research
Scientific Computing Center's T3E supercomputer and we thank them for a
generous allocation of computing time. We would also like to thank Bill
Saphir and the NERSC PC Cluster Project for additional computational
time. The computing support was made possible by the Office of Science
of the U.S. Department of Energy under Contract No. DE-AC03-76SF00098.

\end{document}